\documentclass[aps,superscriptaddress,floats,twocolumn,epsf,prl]{revtex4-2}

\usepackage{graphicx}

\usepackage[dvipsnames]{xcolor}
\usepackage{MnSymbol}
\usepackage{nicefrac}
\usepackage{braket}
\usepackage{array}
\usepackage{bm}
\usepackage{hyperref}
\usepackage{changes}
\usepackage[normalem]{ulem}

\begin{document}

\title{Thermodynamic Stability at the Two-Particle Level}
\author{A.~Kowalski}
\altaffiliation{These authors contributed equally to this work.}
\affiliation{Institut f\"ur Theoretische Physik und Astrophysik and W\"urzburg-Dresden Cluster of Excellence ct.qmat, Universit\"at W\"urzburg, 97074 W\"urzburg, Germany}
\author{M.~Reitner}
\altaffiliation{These authors contributed equally to this work.}
\affiliation{Institute of Solid State Physics, TU Wien, 1040 Vienna, Austria}
\author{L.~Del~Re}
\altaffiliation{These authors contributed equally to this work.}
\affiliation{Department of Physics, Georgetown University, 37th and O Sts., NW, Washington, DC 20057, USA}
\affiliation{Max-Planck-Institut f\"ur Festk\"orperforschung, Heisenbergstra{\ss}e 1, 70569 Stuttgart, Germany}
\author{M.~Chatzieleftheriou}
\affiliation{Laboratoire de Physique et Etude des Mat\'eriaux, UMR8213 CNRS/ESPCI/UPMC, Paris, France}
\affiliation{CPHT, CNRS, \'Ecole polytechnique, Institut Polytechnique de Paris, 91120 Palaiseau, France}
\author{A.~Amaricci}
\affiliation{CNR-IOM DEMOCRITOS, Istituto Officina dei Materiali,
Consiglio Nazionale delle Ricerche, Via Bonomea 265, I-34136 Trieste, Italy}
\author{A.~Toschi}
\affiliation{Institute of Solid State Physics, TU Wien, 1040 Vienna, Austria}
\author{L.~de'~Medici}
\affiliation{Laboratoire de Physique et Etude des Mat\'eriaux, UMR8213 CNRS/ESPCI/UPMC, Paris, France}
\author{G.~Sangiovanni}
\affiliation{Institut f\"ur Theoretische Physik und Astrophysik and W\"urzburg-Dresden Cluster of Excellence ct.qmat, Universit\"at W\"urzburg, 97074 W\"urzburg, Germany}
\author{T. Sch\"afer}
\email{t.schaefer@fkf.mpg.de}
\affiliation{Max-Planck-Institut f\"ur Festk\"orperforschung, Heisenbergstra{\ss}e 1, 70569 Stuttgart, Germany}

\date{ \today }

\begin{abstract}
We show how the stability conditions for a system of interacting fermions that conventionally involve variations of thermodynamic potentials can be rewritten in terms of one- and two-particle correlators. 
We illustrate the applicability of this alternative formulation in a multi-orbital model of strongly correlated electrons at finite temperatures, inspecting the lowest eigenvalues of the generalized charge susceptibility in proximity of the phase-separation region. 
Additionally to the conventional unstable branches, we address unstable solutions possessing a positive, rather than negative compressibility. 
Our stability conditions require no derivative of free energy functions with conceptual and practical advantages for actual calculations and offer a clear-cut criterion for analyzing the thermodynamics of correlated complex systems. 
\end{abstract}

\maketitle

\vskip 5mm

\noindent
{\sl Introduction}. Thermodynamic stability  is a crucial concept for condensed matter systems. The conventional formulation of  stability criteria relies on derivatives of thermodynamic potentials, i.e., on the Hessian matrix of the grand potential $\Omega$, taken with respect to the independent variables considered, such as temperature, volume and chemical potential. 
In the textbook example of the liquid-gas transition \cite{Goldenfeld}, the stability of the van der Waals isotherms in the pressure-{\it vs}-volume plane can be directly inspected by calculating the isothermal compressibility, which can be expressed with the second derivative of the grand potential w.r.t. the volume. 
Similar considerations extend also to many-electron systems in the presence of a local Coulomb interaction. 
The latter induces metal-to-Mott insulator transitions at finite doping, close to which two derivatives are needed in order to set the stability conditions at a fixed temperature: one with respect to the strength of the Hubbard repulsion and one with respect to the chemical potential \cite{Kotliar2000,Tong2001,Kotliar2002,Eckstein2007,Chatzieleftheriou2023}.
The dimension of the Hessian matrix would further increase as one keeps adding thermodynamic variables, hence leading to a higher number of independent derivatives to be considered. One can therefore ask whether it is possible to encode the same information  in a single ``local" state variable, whose very value diagnoses the thermodynamic stability of a system. Extending this concept to multidimensional abstract spaces, such a condition would offer the additional advantage of not having to explore derivatives in all different directions, when those are hard to compute.

\begin{figure}[b]
\includegraphics[width=0.9\columnwidth]{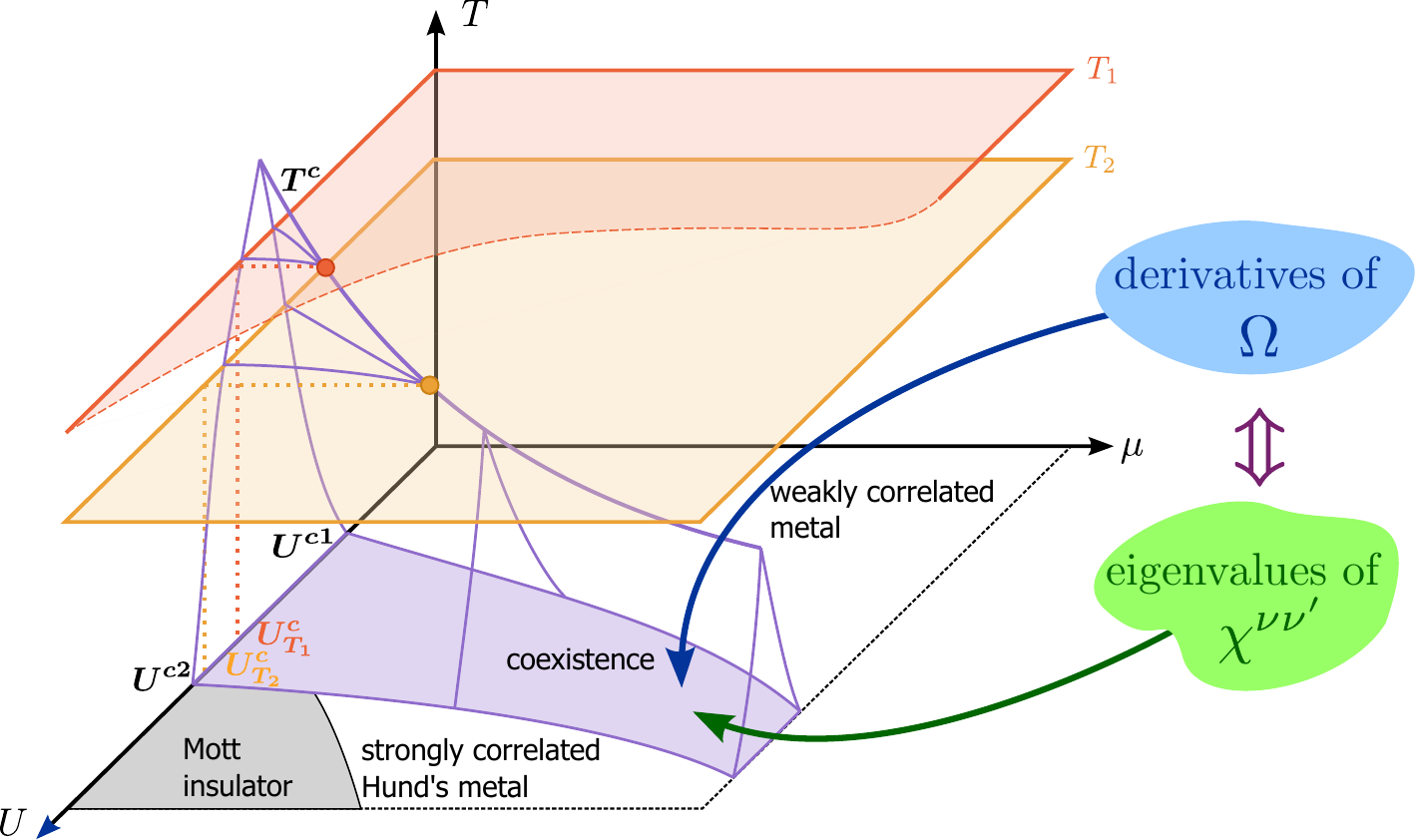}
\caption{Sketch of the phase diagram of the two-orbital Hubbard model described in Eq.~(\ref{eq:Hubbard_2orbs}) as a function of temperature $T$, interaction $U$, and chemical potential $\mu$. The $U$-axis is located at half-filling and $T\!=\!0$. At finite doping, the violet ``moustache"-shaped region describes a coexistence regime of a weakly correlated and a strongly correlated (Hund's) metal. }
\label{fig:sketch}
\end{figure}
\noindent
In this Letter, we show that this is indeed possible by calculating eigenvalues of two-particle vertex functions. Even though a direct connection with the thermodynamic potentials is not obvious at first sight, here we demonstrate that the conditions based on the Hessian can be rewritten solely in terms of the single-particle propagator $G$ and the eigenspectrum of the generalized two-particle susceptibility $\chi$ (see right-hand side of the sketch in Fig.~\ref{fig:sketch}). First, we derive the relevant thermodynamic stability criteria in terms of derivatives of the grand potential $\Omega$ for the single-band Hubbard model. Second, we make the connection to two-particle response functions $\chi$, and show that the stability conditions can be rewritten in terms of their eigenvalues and eigenvectors, giving a clear-cut and compact criterion for the thermodynamic stability of a strongly correlated system. In doing so we also generalize our findings to generic two-body interactions as well as for the case of the presence of a magnetic field \cite{Supplemental}. Eventually, we exemplify the validity of our criterion for a more general system, a multi-orbital Hubbard model: we numerically explore the coexistence region in its phase diagram with chemical potential $\mu$, temperature $T$ and  interaction strength $U$, where the thermodynamic analysis is relevant to assess the stability of the various competing phases (see Fig.~\ref{fig:sketch}).

\noindent
{\sl Thermodynamic stability criteria for the Hubbard model.}
We first focus on the single-band Hubbard model \cite{Hubbard1963, Hubbard1964, Gutzwiller1963, Kanamori1963,Qin2022,Arovas2022}
\begin{equation}
    H\!=\!\sum\limits_{ij\sigma}t_{ij}c^\dagger_{i\sigma}c_{j\sigma}+U\sum\limits_{i}n_{i\uparrow}n_{i\downarrow}-\mu\sum\limits_{i\sigma}n_{i\sigma},
    \label{eq:Hubbard}
\end{equation}
where $t_{ij}$ are hopping amplitudes, $c^{(\dagger)}_{i\sigma}$ annihilates (creates) an electron with spin $\sigma$ on site $i$, $n_{i\sigma}$ is the number operator, $U$ the purely local Coulomb interaction and $\mu$ the chemical potential. For this model,
the grand potential (Landau free energy) $\Omega=-T\ln{Z}$ (with the partition function $Z$) is a function of the free parameters $(T,\mu,U)$ and its exact differential reads~\cite{Sordi2011, Walsh2019}
\begin{equation}
    \frac{1}{V}d\Omega\!=\!-sdT-nd\mu+DdU,
\end{equation}
where $\frac{1}{V}\frac{\partial\Omega}{\partial U}\!=\!D \!=\! \langle n_\uparrow n_\downarrow \rangle$ is the double occupancy, $-\frac{1}{V}\frac{\partial\Omega}{\partial\mu}\!=\!n$ the density, and $-\frac{1}{V}\frac{\partial\Omega}{\partial T}\!=\!s$ the entropy per lattice site at fixed volume $V$. The stability of the solutions requires its Hessian to be negative $d^2\Omega\!<\!0$ \cite{Kotliar1999,Werner2005,Strand2011,Sordi2011, Vanloon2020, Supplemental}. 
If we fix the temperature $T$, this explicitly means that:
\begin{equation}
    d^2\Omega\!=\!\begin{pmatrix}d\mu & dU\end{pmatrix}
    \begin{pmatrix}\frac{\partial^2 \Omega}{\partial\mu^2} &  \frac{\partial^2 \Omega}{\partial U\partial\mu}\\ \frac{\partial^2 \Omega}{\partial U\partial\mu} &  \frac{\partial^2 \Omega}{\partial U^2}\end{pmatrix}
    \begin{pmatrix}d\mu \\ dU\end{pmatrix}\!<\!0.
\end{equation}
The Hessian is negative definite (and, hence, the system is stable) if its principle minors are alternating in sign, beginning with a negative one. This leads to the following general stability criteria:
\begin{eqnarray}
    \frac{\partial^2 \Omega}{\partial\mu^2}&<&0,\\
    \frac{\partial^2\Omega}{\partial\mu^2}\frac{\partial^2\Omega}{\partial U^2}-\left[\frac{\partial^2\Omega}{\partial U \partial\mu}\right]^2&>&0.
\end{eqnarray}
Reexpressed in thermodynamic observables and parameters, these conditions read:
\begin{eqnarray}
    \frac{\partial n}{\partial\mu}&>&0\label{eq:stability_1},\\
    -\frac{\partial n}{\partial\mu}\frac{\partial D}{\partial U} - \left[\frac{\partial n}{\partial U}\right]^2&>&0
    \label{eq:stability_2}
\end{eqnarray}
($-\frac{\partial n}{\partial U} = \frac{\partial D}{\partial \mu}$ holds as a Maxwell relation).
Note that the first condition Eq.~(\ref{eq:stability_1}) is equivalent to the well-known criterion that in a thermodynamically stable system, the electronic compressibility has to be positive, i.e., $\kappa\!=\!\frac{2}{n^2}\frac{\partial n}{\partial \mu}\!>\!0$. However, Eq.~(\ref{eq:stability_1}) is not a sufficient criterion, so that for a system to be thermodynamically stable, also  Eq.~(\ref{eq:stability_2}) has to hold \footnote{We note in passing that if $\kappa\!>\!0$, the following (in-)equality holds: $0\!<\! -\left.\frac{\partial D}{\partial U}\right|_{n} \leq -\left.\frac{\partial D}{\partial U}\right|_{\mu}$ \cite{Schueler2018}.}.
\begin{figure*}[t!]
\includegraphics[width=\textwidth]{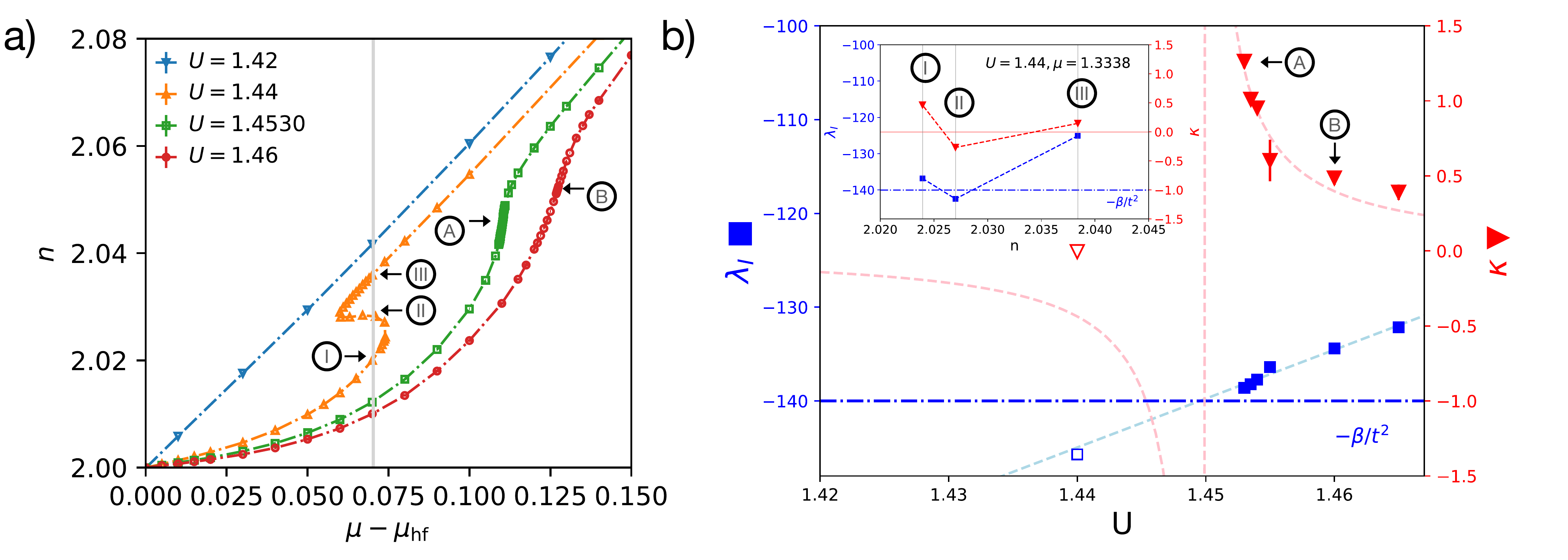}
\caption{DMFT calculations of the two-orbital Hubbard model on the Bethe lattice at $T_1\!=\!1/35$.  a): $n$ vs. $\mu-\mu_\text{hf}$ for several values of the interaction $U$. \textcircled{{\scriptsize A}} and \textcircled{{\scriptsize B}} indicate the respective maxima in $\kappa$ for two different interaction values. The vertical grey line indicates the cut at constant chemical potential $\mu=1.3338$ corresponding to the inset in panel b), \textcircled{{\footnotesize I}}-\textcircled{{\tiny III}} indicate the respective points for $U=1.44$. b): Lowest eigenvalue $\lambda_{I}$ of the generalized static local charge susceptibility $\chi^{\nu\nu'}$ (blue squares) and the corresponding values of the charge susceptibility $\kappa$ (red triangles). Here, the values of the respective maxima of $\kappa$ are shown. The dashed grey lines are fits of the values of $\kappa$ (Eq.~\ref{eq:kappa}) and $\lambda_I$ (linear), respectively, and serve as guide to the eye. Inset: Cut at constant interaction $U=1.44$ and chemical potential [corresponding to the solid grey line in a)].}
\label{fig:beta35}
\end{figure*}

\noindent
{\it Connection to two-particle response functions.}
After deriving the above stability criteria, let us now relate them to the structure of two-particle response functions. In particular, the momentum-dependent static charge susceptibility of the system is given by $\chi(\mathbf{q})\!=\! \tfrac{1}{2}\int_0^\beta d\tau \left<n(\mathbf{q},\tau)n(\mathbf{-q},0)\right>-\left<n\right>^2$, where $\beta\!=\!1/(k_\text{B}T)$ and $n(\mathbf{q})\!=\!\frac{1}{V}\sum_{\mathbf{k}\sigma}c^\dagger_{\mathbf{k+q}\sigma}c_\mathbf{k\sigma}$. Then the corresponding local susceptibility $\chi_\text{loc}$ can be obtained by summing over all momenta $\mathbf{q}$.

The latter response function can be either directly measured by applying a local external field coupled to the charge degrees of freedom or obtained by summing the so-called (local) generalized susceptibility $\chi^{\nu\nu'}$ \cite{Supplemental,Schaefer2013,Reitner2020,Chalupa2021,Pelz2023,Adler2024,Essl2024} at zero bosonic transfer frequency $\Omega=0$ over the two fermionic Matsubara frequencies $\nu$ and $\nu'$:
\begin{eqnarray}
    \chi_\text{loc}\!=\!\frac{1}{\beta^2}\sum\limits_{\nu\nu'}{\chi^{\nu\nu'}(\Omega\!=\!0)}\!=\!\sum\limits_\alpha \lambda_\alpha w_\alpha.
    \label{eq:gen_susc}
\end{eqnarray}
The last equality represents the decomposition in the eigenbasis of $\chi^{\nu\nu'}(\Omega\!=\!0)$ with eigenvalues $\lambda_\alpha$ and eigenvectors $V_\alpha$ constituting the respective spectral weights $w_\alpha\!=\!\left[\tfrac{1}{\beta}\sum_\nu V^{-1}_\alpha(\nu)\right]\left[\tfrac{1}{\beta}\sum_{\nu'} V_\alpha(\nu')\right]$ \cite{Schaefer2016,Reitner2020}. 
The connection between the local susceptibility and the electronic compressibility becomes particularly transparent within the dynamical mean-field theory (DMFT, \cite{Georges1996,Metzner1989,Georges1992,VanLoon2015,VanLoon2024secondorder}), which is the exact solution of Eq.~(\ref{eq:Hubbard}) in the limit of infinite lattice connectivity. 
In the case of the Bethe lattice non-local correlation functions can be analytically expressed \cite{Georges1996,Reitner2020,delre_and_g2021,Vanloon2022} in terms of purely local ones \cite{Rohringer2012,Gunnarsson2017,Chalupa2021,Adler2024}. In particular, the following expression for the compressibility $\kappa=\frac{2}{n^2}\chi(\mathbf{q}\!=\!0)$ holds:
\begin{eqnarray}
    \kappa\!=\!\frac{2}{n^2\beta^2}\sum\limits_{\nu\nu'}\left[\chi^{-1}_{\nu\nu'}+\frac{t^2}{\beta}\delta_{\nu\nu'} \! \right]^{-1} \!=\!\frac{2}{n^2}\sum_\alpha\left[\frac{1}{\lambda_\alpha} \! + \! \frac{ t^2}{\beta}\right]^{-1}  \kern-1em  \,
    w_\alpha.
    \label{eq:kappa}
\end{eqnarray}
Here one can immediately recognize that the first stability criterion, Eq.~(\ref{eq:stability_1}), namely that $\kappa$ should be positive, is intimately related to the spectrum of the generalized local charge susceptibility. This is quite remarkable, since $\kappa$ is a lattice quantity, whereas $\chi^{\nu\nu'}$ represents a purely local impurity quantity. Actually, as we derive in detail in \cite{Supplemental}, {\it both} stability criteria Eqs.~(\ref{eq:stability_1}) and (\ref{eq:stability_2}) depend directly on the eigenvalues $\lambda_\alpha$ and (via accompanying weights $w_\alpha, v_\alpha, y_\alpha$) on the eigenvectors $V_\alpha$ of $\chi^{\nu\nu'}$:
\begin{eqnarray}
    \kappa\!=\!\frac{2}{n^2}\sum_\alpha\left[\frac{1}{\lambda_\alpha}+\frac{t^2}{\beta}\right]^{-1} w_\alpha&>&0\label{eq:stability_1_2p},\\
    -n^2\kappa \left[-\frac{D}{U}+\sum\limits_\alpha\frac{1}{\frac{1}{\lambda_\alpha}+\frac{t^2}{\beta}}v_\alpha\right]\nonumber\hspace{1.5cm}& &\\-\left[\sum\limits_\alpha \frac{2}{\frac{1}{\lambda_\alpha}+\frac{t^2}{\beta}}y_\alpha\right]^2&>&0,\label{eq:stability_2_2p}
\end{eqnarray}
where $y_\alpha =\left(\frac{1}{\beta}\sum_{\nu}V_\alpha(\nu)\right)\left(-\frac{1}{\beta}\sum_{\nu}V^{-1}_{\alpha}(\nu) \left.\frac{\partial \Sigma}{\partial U}\right|_G \right)$, and $v_\alpha = \left(\frac{1}{\beta}\sum_\nu V_\alpha(\nu)\frac{\tilde{\chi}_0^{\nu}}{\chi_0^{\nu}}\right)\left(-\frac{1}{\beta}\sum_\nu V^{-1}_\alpha(\nu)\left.\frac{\partial \Sigma}{\partial U}\right|_G\right)$, with $\left.\frac{\partial \Sigma}{\partial U}\right|_G =   \frac{1}{2U}\left(\frac{1}{\beta}\sum_{\nu^{\prime\prime}}\Gamma_d^{\nu\nu^{\prime\prime}}G_{\nu^{\prime\prime}} + \Sigma_{\nu}\right)$ being the explicit derivative of the self-energy w.r.t. the interaction fixing the Green's function, the lattice bubble $\chi_0^{\nu}(\mathbf{q}\!=\!0)\!=\!-\frac{\beta}{V}\sum_{\mathbf{k}}G^2_k$, and the expression $\tilde{\chi}_0^{\nu}(\mathbf{q}\!=\!0)\!=\!-\frac{\beta}{V\,U}\sum_{\mathbf{k}}G^2_k[i\nu+\mu-\epsilon_{\mathbf{k}}]$.

\noindent
Thus, we have expressed all thermodynamic derivatives given in Eqs.~(\ref{eq:stability_1}) and (\ref{eq:stability_2}) by single and two-particle correlation functions evaluated for a given parameter set. This result, which applies even to the exact solution of the Hubbard model at finite lattice connectivity, is particularly remarkable since one would have naively expected that their determination had also required the knowledge of higher order correlation functions \cite{Supplemental}.
\noindent
This way we managed to translate conditions on thermodynamic derivatives into conditions on the eigenspectrum of $\chi$. In particular, if the value of the lowest eigenvalue $\lambda_I$ falls below the lower bound $-\beta/t^2$\footnote{Note that negative eigenvalues of the generalized susceptibility have  been observed in various systems exhibiting local moment formation~\cite{Gunnarsson2016, Thunstroem2018, Chalupa2021, Adler2024, Essl2024} and also in cluster DMFT calculations of the Hubbard model  \cite{Gunnarsson2016,Vucicevic2018}.}, $\kappa$ in Eq.~(\ref{eq:stability_1_2p}) turns negative, hence signalling an instability of the system. Violations of the second condition Eq.~(\ref{eq:stability_2_2p})  are also dictated by the eigenvalues and eigenvectors of $\chi$. It is important to note that such a  connection between derivatives of thermodynamic potentials and the eigenspectrum of the generalized susceptibility can be readily extended to the case with more than one orbital, as well as to general two-body interactions, finite dimensions and the case of the presence of a magnetic field \cite{Supplemental}.

\noindent
{\sl Stability analysis close to a Mott transition.}
To show the validity of our reformulated thermodynamic analysis, we 
investigate a many-body Hamiltonian which possesses an extended region of instability in the proximity to a Mott transition. In general, the parameter ranges where such an instability occurs is larger in multi-orbital models than in single-band ones \cite{Chatzieleftheriou2020}. For this reason, we address a two-orbital Hubbard model (see, e.g., \cite{Steinbauer2019,Chatzieleftheriou2023}) and consider again the Bethe lattice. We use DMFT to study the paramagnetic phase and consider an interaction of density-density Hubbard form:
\begin{eqnarray}
 H &=&\sum_{\langle i, j \rangle, m, \sigma} t_{ij}  c^{\dag}_{im\sigma} c_{jm\sigma}
        + U \sum_{im} n_{im\uparrow} n_{im\downarrow}\nonumber\\
        &+& (U - 2J) \sum_{im,m' \neq m} n_{im\uparrow} n_{im'\downarrow}\nonumber\\
        &+& (U - 3J) \sum_{i,m < m', \sigma} n_{im\sigma} n_{im'\sigma}\nonumber\\
        &-& \mu\sum\limits_{im\sigma}n_{im\sigma},
        \label{eq:Hubbard_2orbs}
\end{eqnarray}
with $c^{(\dag)}_{im\sigma}$ annihilating (creating) an electron on lattice site $i$ in orbital $m \in \{1, 2\}$ and with spin $\sigma$, density operators $n_{im\sigma} = c^{\dag}_{im\sigma}c_{im\sigma}$, nearest-neighbor hopping matrix elements $t_{ij}$ and as interaction parameters the on-site same-orbital repulsion $U$ and Hund's exchange coupling $J$ which we fix to $J = U/4$. 

\begin{figure*}[t!]
\includegraphics[width=\textwidth]{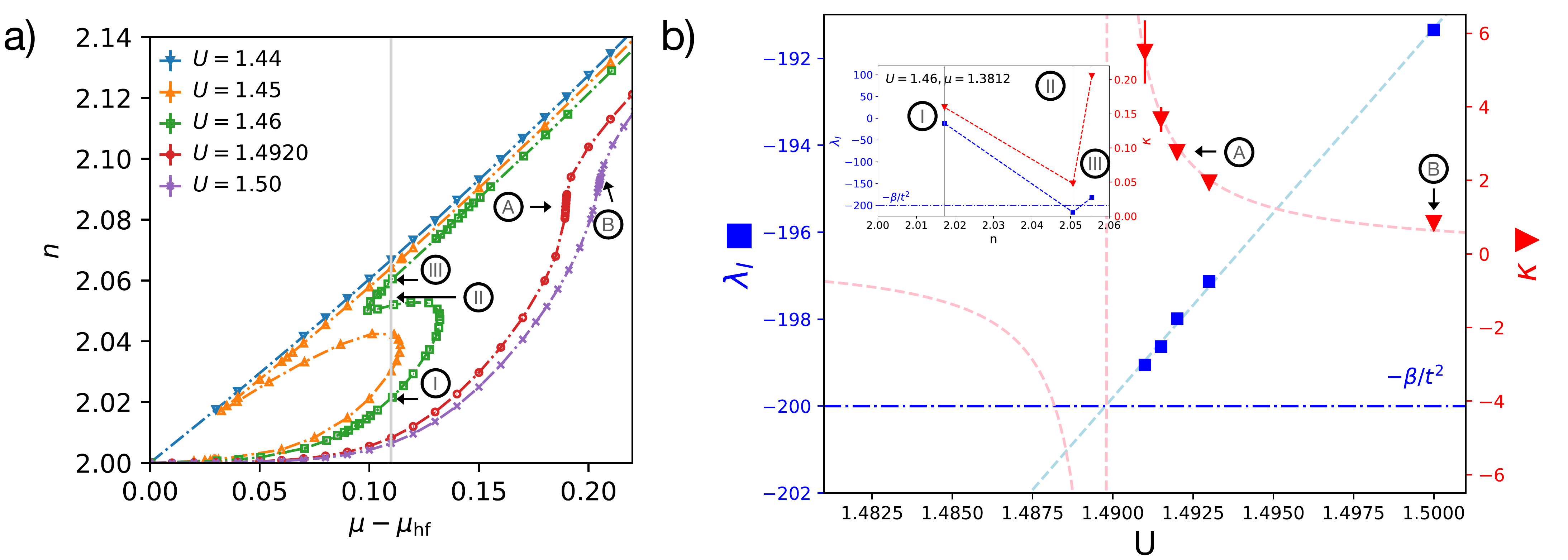}
\caption{Analogous plot to Fig.~\ref{fig:beta35} for $T_2\!=\!1/50$.}
\label{fig:beta50}
\end{figure*}

\noindent
As sketched in the phase diagram as a function of $(T,\mu,U)$ of Fig.~\ref{fig:sketch}, the system undergoes a first-order transition from a (weakly) correlated metal to a Mott insulator at $U^{c2}$ at $T\!=\!0$ and half-filling, upon increasing $U$ \cite{Chatzieleftheriou2023}. Starting from the opposite strong coupling limit, the metallic solution sets in only at $U^{c1}<U^{c2}$ and, hence, a hysteresis region appears for $U^{c1}<U<U^{c2}$. In this region, the equation of state of the system is multi-valued and the insulating solution coexists with the metallic one. At finite doping both solutions become metallic, however they evolve differently when changing the chemical potential $\mu$: the former Mott insulator turns into a strongly correlated Hund's metal \cite{Georges2013}, separated from the weakly correlated metal by a coexistence zone (violet-shaded ``moustache"-shaped). This coexistence regime shows phase separation and is therefore thermodynamically unstable. At finite temperatures, the phase separation region shrinks, terminating at a line of second order critical end points $(T^c,\mu^c,U^c)$. 

\noindent
After these general considerations, let us now illustrate our results, focusing on two specific temperatures $T_1\!=\!1/35$ (Fig.~\ref{fig:beta35}) and $T_2\!=\!1/50$ (Fig.~\ref{fig:beta50}). The overall behavior at different temperatures looks quite similar. However, we will show that for the data set we consider at the lowest temperature $T_2$ the system becomes thermodynamically unstable by violating the stability criteria in a qualitatively different way than for the data set at $T_1$. 
We start with the higher value $T_1$. In panel a) of Fig.~\ref{fig:beta35} we show the total electron filling $n\!=\!\sum_{im\sigma}\left<n_{im\sigma}\right>$ as a function of the chemical potential (measured from the chemical potential at half-filling) for several values of $U$. On the metallic branch the filling depends approximately linearly on the chemical potential, which is the case for the line of $U\!=\!1.42$ \footnote{Note that for interaction strengths above the region of phase separation, this approximately linear dependence holds only for sufficiently high $n$. At lower $n$, we first find a convex section of the occupation approximating a Mott plateau at half-filling followed by a rapid rise that then smoothly crosses over into a concave section which approximates a linear behavior at higher $n$ further away from half-filling.}. Here, the electronic compressibility $\kappa\!=\!1/n^2~\partial n/\partial \mu$ assumes moderate values ($\kappa\approx 0.15$).

\noindent
Approaching the region of phase separation from larger interaction values, we observe a dramatic sharpening of this crossover: At $U=1.46$ we clearly see the onset of a stretched s-shaped curve, which assumes its maximum value of $\kappa\!\approx\!0.5$ at point \textcircled{{\scriptsize B}}. A small decrease of the interaction to $U\!=\!1.453$ results in an even larger compressibility of $\kappa\!\approx\!1.25$ before it eventually diverges, giving rise to the  ``distorted" s-shaped curve of $U\!=\!1.44$.
Tracing the $(n,\mu)$-curve at this particular value of the interaction reveals three distinct regimes (of which we mark three representative points): a strongly correlated metallic regime \textcircled{{\footnotesize I}} and a weakly correlated metallic regime \textcircled{{\tiny III}}, connected by an unstable solution \textcircled{{\tiny II}}. At the boundaries of the stable branches, the compressibility diverges. The (discontinuous) jump from one stable solution to the other would correspond to a Maxwell construction \cite{Chatzieleftheriou2023}.

\noindent
We now more closely inspect the electronic compressibility and its connection to the eigenvalue structure of $\chi^{\nu\nu'}$. Panel b) of Fig.~\ref{fig:beta35} shows the maximum values of $\kappa$ (red triangles) as well as the value of the smallest (leading) eigenvalue of $\chi^{\nu\nu'}$, $\lambda_I$ (blue squares), for several values of the interaction. With $\lambda_I$ we can understand the increase of $\kappa$ when decreasing $U$, since the leading eigenvalue becomes more negative. Eventually, $\lambda_I$ approaches $-\beta/t^2=-140$, at which value Eq.~(\ref{eq:stability_1}) indicates that $\kappa\rightarrow\infty$ \cite{Reitner2020}. Immediately after the divergence, $\kappa$ assumes a negative value, thus violating the first stability condition Eq.~(\ref{eq:stability_1}). To emphasize this violation, in the inset of Fig.~{\ref{fig:beta35} b) we show $\lambda_I$ and $\kappa$ as a function of the total filling for the points \textcircled{{\footnotesize I}}-\textcircled{{\tiny III}} at fixed $U=1.44$ and $\mu=1.3338$ [corresponding to the gray vertical line in panel a)]. One can see that, in this case, for the two stable branches, $\kappa>0$ and $\lambda_I>-\beta/t^2$, whereas for the unstable branch $\kappa<0$ and $\lambda_I\approx -142<-\beta/t^2$.

\noindent
Calculations at the lower temperature $T_2=1/50$ demonstrate that the system can become thermodynamically unstable even if the compressibility is positive.
This is shown in Fig.~\ref{fig:beta50}:
In panel a) we can see that the overall behavior of $\left<n\right>$ as a function of $\mu$ is very similar to the one at elevated temperatures (linear behavior away from the coexistence region, s-shape coming close to it). Additionally, branches connecting the two different metallic regimes are again appearing. In the case of $U\!=\!1.46$ we can see that this {\it clearly unstable} branch contains a regime (e.g., point \textcircled{{\tiny II}}) with \textit{positive} compressibility. Repeating our analysis of the leading eigenvalues $\lambda_I$ in panel b), we see that, also for $T_2$, when $\lambda_I$ is approaching the limit $-\beta/t^2=-200$, the compressibility diverges $\kappa\rightarrow\infty$. The inset again shows that, for $U=1.46$, the leading eigenvalue is $\lambda_I\approx -216 < -\beta/t^2$ for the point on the unstable branch \textcircled{{\tiny II}}. However, since the eigenvalue is further away from the respective limit than in the case of $T_1$ in Fig.~\ref{fig:beta35}, $\kappa$ remains positive.

\noindent
{\it A single criterion for thermodynamic (in-)stability.}
The above numerical findings indeed demonstrate the significance of the eigenvalues of the generalized charge susceptibility for the thermodynamic stability of a strongly correlated system: if one of the eigenvalues falls below the limit $-\beta/t^2$ the relations Eq.~(\ref{eq:stability_1}) and Eq.~(\ref{eq:stability_2}) are not fulfilled at the same time, rendering the system thermodynamically unstable. The main player here can be identified with the lowest eigenvalue $\lambda_I$~\cite{Supplemental}. 
\noindent
Inspecting the structure of the condition based on the eigenspectrum of $\chi$ more closely, one could argue that Eq.~(\ref{eq:stability_2}) is always violated close enough to the Mott metal-insulator transition at half-filling. Indeed, $\kappa \propto \partial n/\partial \mu$ is strongly suppressed there and the first term in Eq.~(\ref{eq:stability_2}) could potentially become smaller than the second one. However, analyzing Eq.~(\ref{eq:stability_2_2p}) tells us that in this situation both terms of the equation vanish with the same power law, and our numerical calculations show that the first term remains larger.
The expression of the thermodynamic derivatives in terms of eigenvalues and eigenweights also explains why neither $\frac{\partial n}{\partial \mu}$ nor $\frac{\partial n}{\partial U}$ diverge at the critical endpoint at half-filling, at odds with $\frac{\partial D}{\partial U}$~\cite{Kotliar2002, Vanloon2020,Reitner2020}.
Thereby, we have also demonstrated a direct connection between $\lambda_I\to -\beta/t^2$ and $\partial D/\partial U \to \infty$ at the critical endpoint at half-filling~\cite{Kotliar2000}.

\noindent
{\sl Conclusions.}
We have derived stability conditions for correlated fermionic matter at the one- and two-particle diagrammatic level by relating the conventional criteria involving derivatives of the grand potential to the eigenvalue structure of the generalized susceptibility \cite{Schaefer2013,Schaefer2016,Gunnarsson2017}. We illustrated the applicability of our reformulation with the example of the coexistence region emerging in the proximity of the Hund's Mott metal-insulator transition in a two-band Hubbard model. There, in addition to the unstable solution with negative compressibility, we were able to identify an unstable branch with \textit{positive} compressibility. In this context we demonstrated that the eigenspectrum of the generalized susceptibility 
represents a clear-cut indicator for thermodynamic stability, conceptually changing the conventional viewpoint based on the evaluation of thermodynamic derivatives.

\begin{acknowledgments}
\noindent
{\sl Acknowledgements.}
The authors are grateful for fruitful discussions with Patrick Chalupa-Gantner, Sergio Ciuchi, M{\'a}rio Malcolms de Oliveira, Dirk Manske, Michael Meixner, Walter Metzner, and Erik van Loon. We also thank Daniel Springer for providing data of the one-orbital model for consistency checks at half-filling. We thank the computing service facility of the MPI-FKF for their support. AK and GS were supported by the Deutsche Forschungsgemeinschaft (DFG, German Research Foundation) through Project-ID 258499086 – SFB 1170 and through the W\"urzburg-Dresden Cluster of Excellence on Complexity and Topology in Quantum Matter ct.qmat (Project-ID 390858490, EXC 2147) and gratefully acknowledge the Gauss Centre for Supercomputing e.\ V.\ (www.gauss-centre.eu) for funding this project by providing computing time on the GCS Supercomputer SuperMUC-NG at Leibniz Supercomputing Centre (www.lrz.de). M.R. acknowledges support as a recipient of a DOC fellowship of the Austrian Academy of Sciences. M. R. acknowledges support of a DOC fellowship of the Austrian Academy of Sciences and from the Austrian Science Fund (FWF) within the project I 5487. A. T. and G. S. also acknowledge financial support from  FWF (I 5868) and  DFG (Nr. 449872909), respectively, projects P1 and P5 of the QUAST research unit (FOR 5249) of the DFG. M.Ch., and L.d.M. are supported by the European Commission through the ERC-CoG2016, StrongCoPhy4Energy, gran agreement No. 724177.
\end{acknowledgments}

\bibliographystyle{unsrturl}
\bibliography{references.bib}
\end{document}


\title{Thermodynamic Stability at the Two-Particle Level\\
{\it -- Supplemental Material --}}

\author{A.~Kowalski}
\altaffiliation{These authors contributed equally to this work.}
\author{M.~Reitner}
\altaffiliation{These authors contributed equally to this work.}
\author{L.~Del~Re}
\altaffiliation{These authors contributed equally to this work.}
\author{M.~Chatzieleftheriou}
\author{A.~Amaricci}
\author{A.~Toschi}
\author{L.~de'~Medici}
\author{G.~Sangiovanni}
\author{T. Sch\"afer}
\email{t.schaefer@fkf.mpg.de}

\maketitle

Within this Supplemental Material we provide details on the definitions and derivations of our thermodynamic stability criteria at the two-particle level. In particular, we provide the thermodynamic stability criteria and derivatives for general non-local interactions and in a magnetic field.

\section{Definition of the generalized susceptibility}
The generalized susceptibility in the one-orbital case is defined as
\begin{eqnarray}
\chi^{\nu\nu'}_{\sigma\sigma'\mathbf{k k'}}(\mathbf{q},\Omega) & = & \int d\tau_1 d\tau_2 d\tau_3 \,
e^{-i\nu\tau_1} e^{i(\nu+\Omega)\tau_2} e^{- i(\nu'+\Omega)\tau_3}  \\ \nonumber
    & \times & \left[ \langle T_\tau c^\dagger_{\mathbf{k}\sigma}(\tau_1) c_{ \mathbf{k+q}\sigma }(\tau_2)
    c^\dagger_{\mathbf{k'+q}\sigma' }(\tau_3) c_{  \mathbf{k'}\sigma'}(0)\rangle  \right. \\ \nonumber
         - \langle &T_\tau& \left.c^\dagger_{ \mathbf{k}\sigma}(\tau_1)
    c_{ \mathbf{k+q}\sigma}(\tau_2) \rangle
     \langle T_\tau   c^\dagger_{  \mathbf{k'+q}\sigma'}(\tau_3) c_{  \mathbf{k'}\sigma'}(0)\rangle \right],
\label{eq:chi}
\end{eqnarray}
where $\sigma, \sigma'$ denote the spins, $\mathbf{k}, \mathbf{k'}, \mathbf{q}$ the momenta, $\nu, \nu'$ are fermionic and $\Omega$ bosonic Matsubara frequencies, $T_\tau$ denotes the imaginary time ordering operator, and $\big< \dots \big> = 1/Z\Tr(\e^{-\beta H} \dots)$ the thermal expectation value. The local generalized susceptibility $\chi^{\nu\nu'}_{\sigma\sigma'}$ is obtained  by summing over all momenta 
\begin{equation}
    \chi^{\nu\nu'}_{\sigma\sigma'}(\Omega) = \frac{1}{V^3}\sum_{\mathbf{k k' q}} \chi^{\nu\nu'}_{\sigma\sigma'\mathbf{k k'}}(\mathbf{q},\Omega),
\end{equation}
which within DMFT and in the limit of infinite lattice connectivity corresponds to the local generalized susceptibility of the auxiliary Anderson impurity model. The respective local generalized susceptibility in the charge channel, we mainly focus on in the main text, is defined as $\chi^{\nu\nu^\prime} = \chi^{\nu\nu^\prime}_{\uparrow\uparrow} + \chi^{\nu\nu^\prime}_{\uparrow\downarrow}$.

For the two-orbital model with orbital indices $m,m'=\{1,2\}$ the local generalized susceptibility (in the longitudinal channel) is defined as
\begin{eqnarray}
\chi^{\nu\nu'}_{\sigma\sigma' m m'}(\Omega) & = & \int d\tau_1 d\tau_2 d\tau_3 \,
e^{-i\nu\tau_1} e^{i(\nu+\Omega)\tau_2} e^{- i(\nu'+\Omega)\tau_3}  \\ \nonumber
    & \times & \left[ \langle T_\tau c^\dagger_{\sigma m}(\tau_1) c_{\sigma m}(\tau_2)
    c^\dagger_{\sigma' m'}(\tau_3) c_{\sigma' m'}(0)\rangle  \right. \\ \nonumber
    &    -  & \left.  \langle T_\tau c^\dagger_{\sigma m}(\tau_1)
    c_{\sigma m}(\tau_2) \rangle
     \langle T_\tau   c^\dagger_{\sigma' m'}(\tau_3) c_{\sigma' m'}(0)\rangle \right].
\label{eq:chi_2orb}
\end{eqnarray}
\section{Thermodynamic stability}
Local thermodynamic stability is ensured for a minimum of the internal energy $E_0$ and thus $d^2E_0>0$~\cite{Landau1980}. The internal energy $E_0$ is a function of the extensive variables $(S,N,{\textstyle \sum_i}\langle \hat{D}_i \rangle)$ of the system~\cite{Tschoegel2000} (in the models of the main text, the volume $V$ is fixed). To obtain $E_0$ we perform the following Legendre transformation of the Landau free energy $\Omega(T,\mu,U)$
\begin{equation}
    dE_0 = d(\Omega + T S + \mu N - U{\textstyle \sum_i} \langle \hat{D}_i \rangle) = TdS + \mu dN - UVdD,
\end{equation}
with $\sum_i \langle \hat{D}_i \rangle = V D$.
Since we have performed Legendre transforms in all parameters in respect to $\Omega$, the Hessian requires $d^2\Omega<0$ for local stability. Note that $E_0$ can be regraded as the energy of the independent particles. By increasing the double occupancy with $\delta D$, the energy of the independent particles $E_0$ reduces by $\delta E_0 = -U V \delta D$. The total energy of the system is obtained from the Legendre transform $dE=d(E_0+U VD)$, with $U\!D$ being the potential energy per lattice site, and reads \cite{Sordi2011}:
\begin{equation}
    dE(S,N,U)\!=\!TdS+\mu dN+VD dU.
    \label{eq:energy}
\end{equation}

\section{Thermodynamic derivatives}
In this section we shall derive the spectral decomposition of the thermodynamic derivatives $\frac{d n}{d U}$ and $\frac{d D}{d U}$ presented in the main text. Let us first note that: $\frac{d n}{dU} = \frac{1}{\beta V}\sum_{k\sigma} \frac{d G_{k\sigma}}{dU}$ and 
\begin{equation}
\frac{dD}{dU} = -\frac{D}{U} + \frac{1}{2U}\frac{1}{V\beta}\sum_{k\sigma}[i\nu +\mu -\epsilon_{\mathbf{k}}]\frac{dG_{k\sigma}}{dU},
\end{equation}
where $k = (\mathbf{k},i\nu)$, by making use of the Galitskii-Migdal formula \cite{Galitskii1958}. Hence, in order to evaluate both quantities, one has to evaluate the  derivative of the Green's function with respect to the Hubbard  $U$ that can be written as:
\begin{align}\label{eq:der0}
\frac{dG_{k\sigma}}{dU} = G^2_{k\sigma}\frac{d\Sigma_{k\sigma}}{dU},
\end{align}
that has been obtained using the Dyson's equation. After a moment of thought, one realizes that the self-energy, which is a functional of the Green's function, depends implicitly on $U$ through $G$ but it also has an explicit dependence on $U$. In fact, we can assume that $\Sigma_{k\sigma} = \sum_{n} U^n \mathcal{K}_n[G]$, where $\mathcal{K}_n[G]$ is a functional of the Green's function containing $2n-1$ fermionic lines. If $n>0$ that could correspond to the skeleton expansion of $\Sigma$ as a function of $G$, however our derivation is more general and still holds even in the case including vanishing~\cite{Rossi2015} and negative values of $n$. Therefore:
\begin{align}\label{eq:der1}
    \frac{d\Sigma_{k\sigma}}{dU} &= \sum_{k^\prime \sigma^\prime}\frac{d\Sigma_{k\sigma}}{dG_{k^\prime \sigma^\prime}}\frac{dG_{k^\prime \sigma^\prime}}{dU} + \left.\frac{\partial \Sigma_{k\sigma}}{\partial U}\right|_{G},
\end{align}
where the last term in the RHS indicates the explicit derivative of $\Sigma$ w.r.t. the Hubbard $U$ fixing the Green's function.
In order to set up a BSE for $\frac{dG}{dU}$ one needs to explicitly evaluate $\left.\frac{\partial \Sigma}{\partial U}\right|_{G}$.
For this task, it is convenient to introduce the following rescaled Green's function ${G}_{k\sigma} = U^{-1/2}\overline{G}_{k\sigma}$.  
The self-energy can be written as a functional of the rescaled quantity as following: $\Sigma[\overline{G}] = U^{1/2}\sum_{n}\mathcal{K}_n[\overline{G}]$. Hence, we have:
\begin{align}\label{eq:der2}
    \frac{d\Sigma_{k\sigma}}{dU} &= \sum_{k^\prime \sigma^\prime}\frac{d\Sigma_{k\sigma}}{d\overline{G}_{k^\prime \sigma^\prime}}\frac{d\overline{G}_{k^\prime \sigma^\prime}}{dU} + \left.\frac{\partial \Sigma_{k\sigma}}{\partial U}\right|_{\overline{G}}. 
\end{align}
The explicit derivative appearing in the last equation can be easily evaluated as $\left.\frac{\partial \Sigma_{k\sigma}}{\partial U}\right|_{\overline{G}}  = \frac{1}{2U}\Sigma_{k\sigma}$. The implicit derivatives appearing in Eq.~(\ref{eq:der2}) can be reformulated in terms of the derivatives appearing in Eq.~(\ref{eq:der1}) by applying the derivative chain rule once again, i.e. 
\begin{align}\label{eq:pass1}
\frac{d\Sigma_{k\sigma}}{d\overline{G}_{k^\prime \sigma^\prime}} = \sum_{k^{\prime \prime}\sigma^{\prime\prime}}\frac{d\Sigma_{k\sigma}}{d{G}_{k^{\prime\prime} \sigma^{\prime\prime}}}\overbrace{\frac{dG_{k^{\prime\prime}\sigma^{\prime\prime}}}{d\overline{G}_{k^\prime \sigma^\prime}}}^{\delta_{k^\prime k^{\prime\prime}}\delta_{\sigma^\prime \sigma^{\prime\prime}}U^{-1/2}} = \frac{U^{-\frac{1}{2}}}{V\beta}\Gamma^{kk^\prime}_{\sigma\sigma^\prime},
\end{align}
and by using the following simple relation:
\begin{align}\label{eq:pass2} 
\frac{d\overline{G}_{k\sigma}}{dU} = \frac{U^{-\frac{1}{2}}}{2}G_{k\sigma} + U^{1/2}\frac{dG_{k\sigma}}{dU}.
\end{align}
In fact, by substituting Eqs.(\ref{eq:pass1},\ref{eq:pass2}) into Eq.~(\ref{eq:der2}), we obtain the following relation:
\begin{align}\label{eq:der3}
    \frac{d\Sigma_{k\sigma}}{dU} &= \frac{1}{V\beta}\sum_{k^\prime \sigma^\prime}\Gamma^{kk^\prime}_{\sigma\sigma^\prime}\frac{dG_{k^\prime \sigma^\prime}}{dU} \nonumber \\ &+ \frac{1}{2U}\left(\frac{1}{V\beta}\sum_{k^\prime\sigma^\prime}\Gamma^{kk^\prime}_{\sigma\sigma^\prime}G_{k^\prime\sigma^\prime} + \Sigma_{k\sigma}\right).
\end{align}
Comparing Eq.~(\ref{eq:der3}) with Eq.~(\ref{eq:der1}) we identify the following useful expression for the explicit derivative of the self-energy:
\begin{align}
    \label{eq:dsigmadU_fixG}
    \left.\frac{\partial\Sigma_{k\sigma}}{\partial U}\right|_G &=\frac{1}{2U}\left(\frac{1}{V\beta}\sum_{k^\prime\sigma^\prime}\Gamma^{kk^\prime}_{\sigma\sigma^\prime}G_{k^\prime\sigma^\prime} + \Sigma_{k\sigma}\right).
\end{align}
Finally, we are able to write the BSE for the Green's function derivative as following:
\begin{align}\label{eq:BSE_vec}
    \frac{dG_{k\sigma}}{dU} &= G^2_{k\sigma} \left.\frac{\partial \Sigma_{k\sigma}}{\partial U}\right|_G+G^2_{k\sigma}\frac{1}{V\beta}\sum_{k^\prime \sigma^\prime}\Gamma^{kk^\prime}_{\sigma\sigma^{\prime}}\frac{dG_{k^\prime \sigma^\prime}}{dU}.
\end{align}
Let us now introduce the generalized susceptibility $\frac{1}{V\beta}\sum_{k^\prime}\chi_{U}^{kk^\prime} = \frac{dG_{k\sigma}}{dU}$, which is a matrix in the quadrimomentum indices. Then the BSE for the generalized susceptibility reads:
\begin{align}\label{eq:BSE_gen}
    \chi_U = \chi_0 \cdot S -\frac{1}{(V\beta)^2}\,\chi_0\cdot \Gamma_d \cdot \chi_U, 
\end{align}
where $\chi_0^{kk^\prime} = -V\beta \delta_{kk^\prime}G_k^2$, $S^{kk^\prime} = -\delta_{kk^\prime}\left.\frac{\partial \Sigma_k}{\partial U}\right|_G$, $\Gamma^{kk^\prime}_d = \Gamma^{kk^\prime}_{\uparrow\uparrow} + \Gamma^{kk^\prime}_{\uparrow\downarrow}$,  $\cdot$ indicates the matrix product, and we used SU(2) symmetry.
After inverting Eq.~(\ref{eq:BSE_gen}) we obtain:
\begin{align}
    \chi_U = \left(\chi_0^{-1} + \frac{1}{(V\beta)^2}\Gamma_d\right)^{-1}\cdot S = \chi_d \cdot S,
\end{align}
where $\chi_d^{kk^\prime}$ is the generalized susceptibility in the charge channel that yields the compressibility once averaged over all the quadrimomenta, i.e., $\kappa = \frac{1}{(V\beta)^2}\sum_{kk^\prime}\chi_d^{kk^\prime}$.
We can now express the derivative of density with respect to the Hubbard interaction as following:
\begin{align}
    \frac{dn}{dU} &= \frac{1}{(V\beta)^2}\sum_{kk^\prime}\chi^{kk^\prime}_{U} = \sum_{\alpha}\lambda_\alpha y_\alpha, 
\end{align}
where $\lambda_{\alpha}$ are the eigenvalues of $\chi^{kk^\prime}_{d}$, that is a matrix in the quadrimomentum indices $k$ and $k^\prime$,  and $y_\alpha = \left(\frac{1}{\beta V}\sum_k V_{\alpha}(k)\right)\left(\frac{1}{V\beta}\sum_k V^{-1}_\alpha(k) S^{kk}\right)$, with $V_{\alpha}(k)$ being the eigenvectors of $\chi^{kk^\prime}_{d}$.

To compute the derivative of the double occupancies w.r.t. the Hubbard U let us multiply both sides of Eq.~(\ref{eq:BSE_gen}) times the diagonal matrix $d^{kk^\prime} = \delta_{kk^\prime}\frac{1}{U}\left(i\nu + \mu - \epsilon_{\mathbf{k}}\right)$, and we obtain the following BSE:
\begin{align}\label{eq:BSE_D}
    \chi_D = d\cdot{\chi}_0\cdot S - \frac{1}{(V\beta)^2}\chi_D\cdot \left(S^{-1}\cdot \Gamma_d\cdot \chi_0\cdot S\right).
\end{align}
where we defined $\chi_D = d\cdot\chi_d\cdot S$.

Analogously for the case of $\frac{dn}{dU}$, it is possible to write the thermodynamic derivatives of double occupations in terms of the spectral decomposition of $\chi_d$ as following:
\begin{align}
    \frac{dD}{dU} &= -\frac{D}{U}+\frac{1}{(\beta V)^2} \sum_{kk^\prime}\chi_D^{kk^\prime} = -\frac{D}{U}+\sum_\alpha \lambda_\alpha v_{\alpha},
\end{align}
where $v_\alpha = \left(\frac{1}{V\beta}\sum_{k}d^{kk}V_{\alpha}(k)\right)\left(\frac{1}{V\beta}\sum_{k}S^{kk}V^{-1}_{\alpha}(k)\right)$. 

\subsection{Expression with higher-order correlation functions}
Alternatively, the expression for the thermodynamic derivatives in the Hubbard model can be obtained straightforwardly by computing derivatives of the partition function. In particular, we have that $\frac{\partial n}{\partial U} = -\frac{1}{V}\frac{\delta^2\log Z}{\delta \mu \delta U}$ and $\frac{\partial D}{\partial U} =\frac{1}{V}\frac{\delta^2\log Z}{\delta U^2}$. Therefore, following this strategy, one would have to compute the following non-trivial quantities:
\begin{align}
    \frac{\delta^2Z}{\delta \mu \delta U} &= -\int_0^\beta d\tau \int_0^\beta  d\tau^\prime\sum_{ij\sigma} \left< T_\tau n_{i\sigma}(\tau) n_{j\uparrow}(\tau^\prime)n_{j\downarrow}(\tau^\prime)\right>, \\
     \frac{\delta^2Z}{\delta U^2} &= \int_0^\beta d\tau \int_0^\beta  d\tau^\prime\sum_{ij} \left<T_\tau n_{i\uparrow}(\tau)n_{i\downarrow}(\tau) n_{j\uparrow}(\tau^\prime)n_{j\downarrow}(\tau^\prime)\right>.
\end{align}
The last two equations are characterized by the presence of three and four body imaginary time-ordered products that are notoriously hard to calculate. On the other hand, Eqs.~(\ref{eq:BSE_gen},\ref{eq:BSE_D}) provide a more efficient way of computing the thermodynamic derivatives that are expressed in terms of on one and two-particle quantities only. 

\subsection{Generalization to non-local interactions}
Eq.~(\ref{eq:dsigmadU_fixG}) and hence Eq.~(\ref{eq:BSE_vec}) can be straightforwardly extended to  non-local interactions of the form:
\begin{equation}
    H_{\mathrm{int}} = \frac{U}{V^3}\sum_{\mathbf{k k' q} 1 2 3 4} \create{\mathbf{k}1} \create{\mathbf{k}'+\mathbf{q} 2} v_{1234}(\mathbf{q},\mathbf{k},\mathbf{k}') \annihil{\mathbf{k}'3} \annihil{\mathbf{k}+\mathbf{q}4},
\end{equation}
where  indices $1, 2, 3, 4$ denote orbital and spin, $U$ gives the intensity of the interaction, and $v_{1234}(\mathbf{q},\mathbf{k},\mathbf{k}')$ the dependence on the momenta and orbital indices. 
The expression for the self-energy $\Sigma_{k12} = \sum_{n} U^{n} \mathcal{K}_n[G]$, that we previously considered is generic and does not depend on the specific form of the interaction. Hence, analogously to the the case of on-site interaction, the following expression for the Green's function derivative holds:
\begin{equation}
\begin{split}
    &\frac{dG_{k12}}{dU} = G^2_{k12} \left.\frac{\partial \Sigma_{k12}}{\partial U}\right|_G+G^2_{k12}\frac{1}{V\beta}\sum_{k^\prime 34}\Gamma^{kk^\prime}_{1234}\frac{dG_{k^\prime34}}{dU}\\
    &=-\frac{1}{V\beta}\sum_{k^\prime 34} \left(\chi^{-1}\right) ^{kk'}_{1234} \left.\frac{\partial \Sigma_{k'34}}{\partial U}\right|_G,
\end{split}
\end{equation}
where:

\begin{align}
    \left.\frac{\partial\Sigma_{k12}}{\partial U}\right|_G &=\frac{1}{2U}\left(\frac{1}{V\beta}\sum_{k^\prime34}\Gamma^{kk^\prime}_{1234}G_{k^\prime34} + \Sigma_{k12}\right),
\end{align}
and 
\begin{equation}
    \chi^{kk'}_{1234} = -\frac{V\beta \delta_{k k'} \delta_{13} \delta_{24}}{G^2_{k12}} + \frac{1}{(V\beta)^2} \Gamma^{kk^\prime}_{1234} 
\end{equation}
is the corresponding generalized susceptibility.

The respective expression for the Landau free energy reads
\begin{equation}
    \frac{1}{V}d\Omega\!=\!-sdT-nd\mu+\mathcal{D}dU,
\end{equation}
where $\mathcal{D} = \frac{1}{UV} \left< H_{\mathrm{int}}\right>$. Local thermodynamic stability is then determined by the derivatives  $\frac{\partial n}{\partial U} = \frac{1}{V\beta}\sum_{k1}\frac{dG_{k11}}{dU}$ and $\frac{\partial \mathcal{D}}{\partial U}$, the latter reads: 
\begin{equation}
    \frac{d \mathcal{D}}{d U}= -\frac{\mathcal{D}}{U}+\frac{1}{U}\frac{1}{V\beta}\sum_{k12} [i\nu + \mu -\epsilon_{\vk1}] \frac{dG_{k12}}{dU}.
\end{equation}

\subsection{Limit of infinite dimensions}
All the results that we have obtained so far hold for the exact solution of the Hubbard model and do not depend on the dimensionality. In the limit of infinite dimensions, where DMFT is exact, further simplifications arise in the expression of the thermodynamic derivatives.  It is convenient to work with the Bethe lattice where it is possible to express analytically homogeneous \cite{Georges1996,delre2018,Reitner2020} and staggered \cite{delre_and_g2021} response functions in terms of local quantities. 
In this case, the expression for the generalised susceptibility in Eq.~(\ref{eq:BSE_gen}) simplifies as following:
\begin{align}
    \chi_U &= \left(\chi^{-1} + \frac{t^2}{\beta}\delta_{\nu\nu^\prime}\right)^{-1}\cdot S^\text{loc},
\end{align}
where $S^{\text{loc}}_{\nu\nu^\prime} =  -\delta_{\nu\nu^\prime}\frac{1}{2U}\left(\frac{1}{\beta}\sum_{\nu^{\prime\prime}}\Gamma_d^{\nu\nu^{\prime\prime}}G_{\nu^{\prime\prime}} + \Sigma_{\nu}\right)$, and $G_\nu = \frac{1}{V}\sum_{\mathbf{k}}G_k$ is the local Green's function. Hence, using the spectral decomposition for $\kappa$ we obtain the following spectral decomposition for $\frac{dn}{dU}$:
\begin{align}
    \frac{dn}{dU} = \frac{2}{\beta^2}\sum_{\nu\nu^\prime}\chi_U^{\nu\nu^\prime} = 2\sum_\alpha \left(\frac{1}{\lambda_\alpha} +\frac{t^2}{\beta}\right)^{-1}y_\alpha, 
\end{align}
where:
\begin{align}
    y_\alpha &= \left(\frac{1}{\beta}\sum_{\nu}V_\alpha(\nu)\right)\left(\frac{1}{\beta}\sum_{\nu}V^{-1}_{\alpha}(\nu)S^{\text{loc}}_{\nu\nu}\right).
\end{align} 
Analogously, we obtain the Bethe-Salpeter equation for $\chi_D$ in infinite dimensions, by first averaging Eq.~(\ref{eq:BSE_D}) over the crystalline momenta and then inverting it, which reads:
\begin{align}
    \chi_D &=  \chi_0^{-1}\cdot \tilde{\chi}_0\cdot\left(\chi^{-1} + \frac{t^2}{\beta}\delta_{\nu\nu^\prime}\right)^{-1}\cdot S^{\text{loc}},
\end{align}
where here $\chi_0^{\nu\nu^\prime} = -\delta_{\nu\nu^\prime}\frac{\beta}{V}\sum_{\mathbf{k}}G^2_k$ and $\tilde{\chi}_0^{\nu\nu^\prime} = -\delta_{\nu\nu^\prime}\frac{\beta}{V\,U}\sum_{\mathbf{k}}G^2_k[i\nu+\mu-\epsilon_{\mathbf{k}}]$, simplifying to 
\begin{equation}
\frac{\tilde{\chi^{\nu\nu}_0}}{\chi^{\nu\nu}_0} = \frac{1}{U} (\ii\nu+\mu-2t^2 G_\nu).
\end{equation}

Hence, the spectral decomposition for $\frac{d{D}}{dU}$  reads:
\begin{align}
    \frac{d{D}}{dU} &= -\frac{D}{U} +\frac{1}{\beta^2}\sum_{\nu\nu^\prime}\chi_D^{\nu\nu^\prime} \nonumber \\
    & = -\frac{D}{U} +\sum_{\alpha}\left(\frac{1}{\lambda_\alpha} +\frac{t^2}{\beta}\right)^{-1} v_\alpha,
\end{align}
where:
\begin{align}
    v_\alpha &= \left(\frac{1}{\beta}\sum_\nu V_\alpha(\nu)\frac{\tilde{\chi}_0^{\nu\nu}}{\chi_0^{\nu\nu}}\right)\left(\frac{1}{\beta}\sum_\nu V^{-1}_\alpha(\nu)S^{\text{loc}}_{\nu\nu}\right).
\end{align}

Let us note, that in the limit of infinite dimensions, the expressions of the thermodynamic derivatives  in terms of the spectral decomposition of $\chi_d$ is a particular case of the generic one that we discussed in the previous section. In particular, in infinite dimensions, it is sufficient to evaluate the spectrum of the local two-particle charge susceptibility for obtaining the derivatives.

\subsection{Stability in a magnetic field}
The thermodynamic stability criteria of the main text can be straightforwardly generalized to case of a system in a static and uniform magnetic field $h$. For simplicity let us focus on the single-band Hubbard model
\begin{equation}
H\!=\!\sum_{\vk\sigma}(\epsilon_\vk-\mu) n_{\vk\sigma} - h \sum_{i}  (n_{i\uparrow}-n_{i\downarrow})+U\sum\limits_{i}n_{i\uparrow}n_{i\downarrow},
\end{equation}
the corresponding  Landau free energy then reads:
\begin{equation}
    \frac{1}{V}d\Omega\!=\!-sdT-nd\mu- m dh+DdU ,
\end{equation}
with $m=\frac{1}{V}\sum_i\left< n_{i\uparrow}-n_{i\downarrow}\right>$. Since $mV$ is an extensive variable, for fixed temperature $T$ the criteria for local stability becomes
\begin{equation}
    d^2\Omega\!=\!\begin{pmatrix}d\mu & dh & dU\end{pmatrix}
    \begin{pmatrix}\frac{\partial^2 \Omega}{\partial\mu^2} & \frac{\partial^2 \Omega}{\partial h\partial\mu}&  \frac{\partial^2 \Omega}{\partial U\partial\mu}\\ 
    \frac{\partial^2 \Omega}{\partial h\partial\mu}& \frac{\partial^2 \Omega}{\partial h^2} & \frac{\partial^2 \Omega}{\partial U\partial h} \\
    \frac{\partial^2 \Omega}{\partial U\partial\mu}& \frac{\partial^2 \Omega}{\partial U\partial h} &  \frac{\partial^2 \Omega}{\partial U^2}\end{pmatrix}
    \begin{pmatrix}d\mu \\ dh \\ dU\end{pmatrix}\!<\!0.
\end{equation}
The Hessian is negative definite if the principle minors are alternating in sign, reexpressed in thermodynamic variables this leads to the following stability criteria:
\begin{align}
    -\frac{\partial n}{\partial \mu}\! &<\!0,\\
    -\frac{\partial n}{\partial \mu}\frac{\partial D}{\partial U} - \left[\frac{\partial n}{\partial U}\right]^2\! &>\!0,\\
     \frac{\partial n}{\partial \mu} \frac{\partial m}{\partial h} \frac{\partial D}{\partial U} - 2 \frac{\partial n}{\partial h}\frac{\partial m}{\partial U}\frac{\partial n}{\partial U} 
     + \frac{\partial n}{\partial \mu}\left[\frac{\partial m}{\partial U}\right]^2\!&  \\\nonumber
      + \frac{\partial m}{\partial h}\left[\frac{\partial n}{\partial U}\right]^2- \frac{\partial D}{\partial U}\left[\frac{\partial n}{\partial h}\right]^2\! &<\!0.
\end{align}
Thereby the submatrix 
\begin{equation}
     \begin{pmatrix}\frac{\partial n}{\partial \mu} &\frac{\partial n}{\partial h}\\ 
    \frac{\partial m}{\partial \mu}& \frac{\partial m}{\partial h} 
    \end{pmatrix} = \frac{2}{(V\beta)^2}\sum_{kk' }  
    \begin{pmatrix}
    \chi^{kk'}_d & \chi^{kk'}_{dm} \\
    \chi^{kk'}_{md} & \chi^{kk'}_{m} 
    \end{pmatrix}
\end{equation}
($\frac{\partial n}{\partial h} = \frac{\partial m}{\partial \mu}$ holds as a Maxwell relation) can be related to the generalized susceptibilities
\begin{align}
\label{eq:sus_d}
     \chi^{kk'}_d &= \frac{1}{2} (\chi^{kk'}_{\uparrow\uparrow}+\chi^{kk'}_{\downarrow\downarrow}+\chi^{kk'}_{\uparrow\downarrow}+\chi^{kk'}_{\downarrow\uparrow})\\
\label{eq:sus_m}
     \chi^{kk'}_m &= \frac{1}{2} (\chi^{kk'}_{\uparrow\uparrow}+\chi^{kk'}_{\downarrow\downarrow}-\chi^{kk'}_{\uparrow\downarrow}-\chi^{kk'}_{\downarrow\uparrow})\\
\label{eq:sus_dm}
     \chi^{kk'}_{dm} &= \frac{1}{2} (\chi^{kk'}_{\uparrow\uparrow}-\chi^{kk'}_{\downarrow\downarrow}-\chi^{kk'}_{\uparrow\downarrow}+\chi^{kk'}_{\downarrow\uparrow})\\
\label{eq:sus_md}
     \chi^{kk'}_{md} &= \frac{1}{2} (\chi^{kk'}_{\uparrow\uparrow}-\chi^{kk'}_{\downarrow\downarrow}+\chi^{kk'}_{\uparrow\downarrow}-\chi^{kk'}_{\downarrow\uparrow})
\end{align}
and the derivatives 
\begin{align}
    \frac{\partial m}{\partial U} &= \frac{1}{V\beta}\sum_{k} \left[\frac{dG_{k\uparrow}}{dU} - \frac{dG_{k\downarrow}}{dU}\right],\\
    \frac{\partial n}{\partial U} &= \frac{1}{V\beta}\sum_{k} \left[\frac{dG_{k\uparrow}}{dU} + \frac{dG_{k\downarrow}}{dU}\right],
\end{align}
can be evaluated with the help of Eq.~(\ref{eq:BSE_gen}) for $\frac{dG_{k\sigma}}{dU}$. $\frac{\partial D}{\partial U}$ can be calculated analogously to the $h=0$ case
\begin{equation}
    \frac{d \mathcal{D}}{d U}= -\frac{\mathcal{D}}{U}+\frac{1}{U}\frac{1}{V\beta}\sum_{k\sigma} [i\nu + \mu -\epsilon_{\vk}] \frac{dG_{k\sigma}}{dU}.
\end{equation}

For the Bethe lattice in infinite dimensions the generalized susceptibilities of the lattice can be reexpressed by the local susceptibilities of the impurity~\cite{Georges1996,DelRe2021,delre_and_g2021}:
\begin{equation}
\begin{aligned}
&\frac{1}{(V\beta)^2}\sum_{kk' } 
     \begin{pmatrix}
    \chi^{kk'}_d & \chi^{kk'}_{dm} \\
    \chi^{kk'}_{md} & \chi^{kk'}_{m} 
    \end{pmatrix} =\\  &\frac{1}{\beta^2}\sum_{\nu\nu' } \left(\begin{pmatrix}
    \chi^{\nu\nu'}_d & \chi^{\nu\nu'}_{dm} \\
    \chi^{\nu\nu'}_{md} & \chi^{\nu\nu'}_{m} 
    \end{pmatrix}^{-1} + \frac{t^2}{\beta}\begin{pmatrix}
    \delta_{\nu\nu'} & 0 \\
    0 & \delta_{\nu\nu'} 
    \end{pmatrix}\right)^{-1},
    \end{aligned}
\end{equation}
where $\chi^{\nu\nu'}_d, \chi^{\nu\nu'}_{dm}, \chi^{\nu\nu'}_{md}, \chi^{\nu\nu'}_{m}$ denote the respective spin combinations of Eqs.~(\ref{eq:sus_d}~-~\ref{eq:sus_md}) for the local generalized susceptibilities.

\section{Landau Functional perspective on thermodynamic stability}
In this section, we alternatively derive the thermodynamic stability expressions of the main text for the case of the Bethe lattice in DMFT by making use of the Landau functional formalism~\cite{Kotliar1999,Kotliar2000} in similar spirit as in Refs.~\cite{Georges1996, Vanloon2020, Reitner2020, Vanloon2022}. Thereby, we can again re-express  $\kappa$, $\partial D/\partial U$, and $\partial n/\partial U$ in terms of the generalized susceptibility $\chi^{\nu \nu'} = \chi^{\nu \nu'}_{\uparrow\uparrow} + \chi^{\nu \nu'}_{\uparrow\downarrow}$:\\

In DMFT the effective action of the auxiliary impurity model reads \cite{Georges1996}
\begin{equation}
    S_{\mathrm{imp}} = - \sum_{\nu \sigma} c^*_{\nu \sigma} \left(\ii\nu + \mu -\Delta_\nu \right) c^{\phantom{*}}_{\nu \sigma} + U \sum_{\omega} n_{\omega \uparrow} n_{-\omega \downarrow},
\end{equation}
where $\Delta_\nu$ is the hybridization function and $c^*_{\nu \sigma}, c^{\phantom{*}}_{\nu \sigma}$ refer to the Fourier transforms of the Grassman variables $c^*_{\sigma}(\tau), c^{\phantom{*}}_{\sigma}(\tau)$:
\begin{eqnarray}
     c^{\phantom{*}}_{\nu \sigma} & = & \frac{1}{\sqrt{\beta}} \int^\beta_0 \dd \tau\,  c^{\phantom{*}}_\sigma(\tau) \e^{\ii \nu \tau} \\
    c^{*}_{\nu  \sigma} & = & \frac{1}{\sqrt{\beta }} \int^\beta_0 \dd \tau\,  c^{*}_\sigma(\tau) \e^{-\ii \nu \tau} \\
    n_{\omega \sigma} & = & \frac{1}{\sqrt{\beta }} \sum_\nu c^*_{\nu \sigma} c_{\nu+\omega \sigma}.
\end{eqnarray}
The corresponding Landau free energy of the impurity is
\begin{eqnarray}
    \Omega_{\mathrm{imp}} & = & -\frac{1}{\beta} \ln{Z_{\mathrm{imp}}} \\
    Z_{\mathrm{imp}} & = & \int \mathcal{D}[c^*,c^{\phantom{*}}]\, \e^{-S_{\mathrm{imp}}}.
\end{eqnarray}
Following Refs.~\cite{Kotliar1999,Kotliar2000,Vanloon2020,Vanloon2022} the DMFT equations for the Bethe lattice can be obtained from an extremum condition of the Landau functional
\begin{equation}
    \Omega[\Delta_\nu]=\Omega_{\mathrm{imp}}[\Delta_\nu] - \frac{1}{\beta}\sum_\nu \frac{\Delta_\nu^2}{t^2}
\end{equation}
by taking the functional derivative with respect to $\delta\Delta_\nu$
\begin{equation}
    \frac{\beta}{2} \frac{\delta \Omega}{\delta \Delta_\nu} = G_\nu - \frac{\Delta_\nu}{t^2} = 0,
\end{equation}
where 
\begin{equation}
G_\nu = \frac{\beta}{2}\frac{\delta \Omega_{\mathrm{imp}}}{\delta \Delta_\nu} = -\frac{1}{2 Z_{\mathrm{imp}}} \frac{\delta Z_{\mathrm{imp}}}{\delta \Delta_\nu}
\end{equation}
refers to the impurity Green's function. The algorithmic stability of the DMFT solution is determined by the eigenvalues $m_\alpha$ of the Hessian $M^{\nu \nu'}$ of $\Omega[\Delta_\nu]$ \cite{Kotliar2000,Vanloon2020,Vanloon2022}
\begin{equation}
\label{eq:hessian}
M^{\nu \nu'}= \frac{\beta}{2} \frac{\delta^2\Omega}{\delta\Delta_\nu\delta\Delta_{\nu'}} = -\frac{1}{\beta}\chi^{\nu \nu'} - \frac{1}{t^2}\delta_{\nu\nu'},
\end{equation}
 which expressed in terms of the eigenvalues $\lambda_\alpha$ of $\chi^{\nu \nu'}$ ($\sum_{\nu'} \chi^{\nu \nu'} V_{\nu'}^\alpha = \lambda_\alpha V_{\nu}^\alpha$  ) are
\begin{equation}
    m_\alpha = -\frac{1}{\beta}\left(\lambda_\alpha +\frac{\beta}{t^2}\right).
\end{equation}
For a stable solution all eigenvalues are negative $m_\alpha<0$, at the critical point one reaches zero $m_\alpha=0$, and for an unstable DMFT solution at least one flips sign\cite{Kotliar2000,Strand2011,Vanloon2020}.

Similarly to the Hessian, also the thermodynamic stability criteria can be related to the generalized susceptibility. Starting with the compressibility, which reads $\kappa=\frac{1}{n^2}\frac{\partial n}{\partial \mu}$, we first consider the derivative of $\frac{\partial G_\nu}{\partial \mu}$~\cite{VanLoon2015} to obtain the change of $n(\mu)=\frac{2}{\beta}\sum_\nu \e^{\ii\nu0^+} G_\nu $
\begin{equation}
    \label{eq:dgdmu}
    \frac{\partial G_\nu}{\partial \mu} = \left. \frac{\partial G_\nu}{\partial \mu} \right|_{\Delta} + \sum_{\nu'} \frac{\delta G_\nu}{\delta \Delta_{\nu'}} \frac{\partial\Delta_{\nu'}}{\partial \mu}.
\end{equation}
For the derivative of the functional expression the implicit dependence of $\Delta_{\nu'}$ on $\mu$ must be thereby considered. Evaluating Eq.~(\ref{eq:dgdmu}) for the DMFT solution at $\Delta_\nu=t^2 G_\nu$  we get 
\begin{eqnarray}
    \frac{\partial G_\nu}{\partial \mu} & = & \frac{1}{\beta}\sum_{\nu'} \left(\chi^{\nu\nu'} -  \chi^{\nu\nu'}  t^2 \frac{\partial G_{\nu'}}{\partial \mu} \right) \\
    \Rightarrow  \frac{\partial G_\nu}{\partial \mu} & = & \frac{1}{\beta}\sum_{\nu'}\left[\chi^{-1}_{\nu\nu'} + \frac{t^2}{\beta} \delta_{\nu\nu'} \right]^{-1},
\end{eqnarray}
which leads to 
\begin{eqnarray}
    \frac{\partial n}{\partial \mu} & = & \frac{2}{\beta^2} \sum_{\nu \nu'} \left[\chi^{-1}_{\nu\nu'} + \frac{t^2}{\beta} \delta_{\nu\nu'} \right]^{-1} \\
    \kappa  & = & \frac{2}{n^2} \sum_{\alpha} \left[\frac{1}{\lambda_\alpha} + \frac{t^2}{\beta} \right]^{-1} w_\alpha
\end{eqnarray}
with 
\begin{equation}
w_\alpha=\left(\frac{1}{\beta}\sum_\nu  V_{\nu}^\alpha\right) \left(\frac{1}{\beta}\sum_{\nu'}  V_{\nu'}^{ \alpha-1}\right).    
\end{equation}
Note that for particle-hole symmetry at half-filling (and due to the SU(2) symmetry of the model), $\chi^{\nu \nu'}$ becomes a bisymmetric matrix~\cite{Springer2020} and the respective eigenvectors $V^\alpha_\nu$, $V^{\alpha-1}_\nu$ are then either symmetric ($V^\alpha_\nu=V^\alpha_{-\nu}$) or antisymmetric ($V^\alpha_\nu=-V^\alpha_{-\nu}$) in $\nu$. There, the eigenvector $V^I_\nu$ of the smallest leading eigenvalue $\lambda_I$ turns out to be antisymmetric, and therefore $w_I=0$. Hence, at half-filling $\lambda_I$ does not contribute to the isothermal compressibility (even if $\lambda_I = -\frac{\beta}{t^2}$) \cite{Vanloon2020,Reitner2020}.
We proceed to calculate $\frac{\partial G_\nu}{\partial U}$\footnote{A somewhat similar derivation for the self-energy is shown in Ref.~\cite{Vanloon2022}.} to obtain $\partial n/\partial U$ and $\partial D/\partial U$:
\begin{equation}
    \label{eq:dgdu}
    \frac{\partial G_\nu}{\partial U} = \left. \frac{\partial G_\nu}{\partial U} \right|_{\Delta} + \sum_{\nu'} \frac{\delta G_\nu}{\delta \Delta_{\nu'}}  \frac{\partial \Delta_{\nu'}}{\partial U}
\end{equation}
Evaluating Eq.~(\ref{eq:dgdu}) for $\Delta_\nu = t^2 G_\nu$ we get
\begin{eqnarray}
    \label{eq:dgdudmft}
    \frac{\partial G_\nu}{\partial U} = \sum_{\nu'} \left[\delta_{\nu \nu'} + \frac{t^2}{\beta}\chi^{\nu\nu'}\right]^{-1} \left. \frac{\partial G_{\nu'}}{\partial U} \right|_{\Delta}.
\end{eqnarray}
Here, $\frac{\partial G_{\nu}}{\partial U}|_{\Delta}$, typically a small antisymmetric function of $\nu$, could be in general calculated from a complicated six-point object. However, the derivative of the auxiliary Anderson impurity Green's function w.r.t.~$U$ for fixed $\Delta_\nu$  can be again related to the explicit derivative of the self-energy $\frac{\partial \Sigma_{\nu}}{\partial U}|_{G}$, which has been introduced in the preceding sections:
\begin{eqnarray}
\left. \frac{\partial G_{\nu}}{\partial U} \right|_{\Delta} & = & G_{\nu}^2 \left(\left. \frac{\partial \Sigma_{\nu}}{\partial U} \right|_{G} + \sum_{\nu'} \frac{\delta \Sigma_{\nu}}{\delta G_{\nu'}} \left. \frac{\partial G_{\nu'}}{\partial U} \right|_{\Delta} \right)\\
\Rightarrow  \left. \frac{\partial G_{\nu}}{\partial U} \right|_{\Delta} & = & - \sum_{\nu'}\left[- \frac{1}{G_{\nu}^2}\delta_{\nu\nu'} + \frac{1}{\beta}\Gamma^{\nu \nu'}_d \right]^{-1} \\
{} & = & - \frac{1}{\beta}\sum_{\nu'} \chi^{\nu \nu'} \left. \frac{\partial \Sigma_{\nu'}}{\partial U} \right|_{G},
\end{eqnarray}
where
\begin{equation}
\left. \frac{\partial \Sigma_{\nu'}}{\partial U} \right|_{G} = \frac{1}{2U}\left(\frac{1}{\beta}\sum_{\nu'}\Gamma^{\nu \nu'}G_{\nu'} + \Sigma_{\nu}\right).
\end{equation}
Hence, by expressing Eq.~(\ref{eq:dgdudmft}) in the eigenbasis of $\chi^{\nu \nu'}$ we get
\begin{equation}
    \frac{\partial n}{\partial U}  =  \sum_\alpha \frac{2}{\frac{1}{\lambda_\alpha}+\frac{t^2 }{\beta}} y_\alpha
\end{equation}
with \begin{equation}
y_\alpha = \left(\frac{1}{\beta}\sum_\nu V^\alpha_\nu \right) \left(-\frac{1}{\beta}\sum_{\nu'} V^{\alpha-1}_{\nu'} \left.\frac{\partial \Sigma_{\nu'}}{\partial U} \right|_{G} \right).
\end{equation}
Again, note that for particle-hole symmetry at half-filling the contribution of the lowest eigenvalue also vanishes $y_I=0$ , since $V^I_\nu$ becomes antisymmetric in $\nu$.
To obtain $\frac{\partial D}{\partial U}$ we make again use of the Galitskii-Migdal formula \cite{Galitskii1958} and obtain
\begin{align}
    \frac{\partial D}{\partial U}  = &  -\frac{D}{U} \\\nonumber + \frac{1}{\beta U} & \sum_{\nu \nu'} (\ii\nu + \mu -2t^2 G_\nu) \left[\delta_{\nu \nu'} + \frac{t^2}{\beta}\chi^{\nu\nu'}\right]^{-1} \left. \frac{\partial G_{\nu'}}{\partial U} \right|_{\Delta}\\
    \frac{\partial D}{\partial U} = &  -\frac{D}{U} +\sum_\alpha\frac{1}{\frac{1}{\lambda_\alpha}+\frac{t^2 }{\beta}}v_\alpha
\end{align}
with 
\begin{align}
v_\alpha = & \frac{1}{U}\left(\frac{1}{\beta}\sum_\nu (\ii\nu + \mu -2t^2 G_\nu) V^\alpha_\nu\ \right)  \\\nonumber
& \times \left(-\frac{1}{\beta}\sum_{\nu'} V^{\alpha-1}_{\nu'} \left.\frac{\partial \Sigma_{\nu'}}{\partial U} \right|_{G}\right)
\end{align}
(where in general, $v_I$ does not vanish for particle-hole symmetry at half-filling).
The obtained relations lead to the thermodynamic stability conditions in the main text.

\section{Generalizations of the thermodynamic stability conditions to the multi-orbital case}
For the 2-orbital model of the main text, the interacting part of the impurity Hamiltonian reads
\begin{equation}
\begin{aligned}
    H_{\mathrm{imp}} = & U (n_{1 \uparrow} n_{1 \downarrow} + n_{2 \uparrow} n_{2 \downarrow}) \\
    &+ (U-2J) (n_{1 \uparrow} n_{2 \downarrow} + n_{2 \uparrow} n_{1 \downarrow}) \\
    &+ (U-3J) (n_{1 \uparrow} n_{2 \uparrow} + n_{1 \downarrow} n_{2 \downarrow}),
\end{aligned}
\end{equation}  
where $J=U/4$. The (renormalized) expectation value of the impurity interaction $\mathcal{D} := \left<H_{imp}\right>/U$ for this model can be seen as a generalization of the double occupancy  $D$ of the one orbital case, and replaces the corresponding quantity in the thermodynamic relations and stability criteria
\begin{equation}
    dE/V = T ds + \mu dn + \mathcal{D} dU
\end{equation}
($s=S/V$).
For the two-orbital case, we again relate the thermodynamic stability conditions with the generalized susceptibility $\chi^{\nu\nu'}$. The effective impurity action of the 2-orbital model in DMFT reads
\begin{equation}
\begin{aligned}
        S_{\mathrm{imp}} =& - \sum_{ \nu \sigma} \Big[
        c^{*}_{\nu 1 \sigma}
            \left(\ii \nu + \mu - \Delta_{\nu 1}\right)
        c_{\nu 1 \sigma} \\
        & \phantom{- \sum_{ \nu \sigma} } +  c^{*}_{\nu 2 \sigma}
        \left(\ii \nu + \mu - \Delta_{\nu 2}\right)
    c_{\nu 2 \sigma} \Big]\\
        &+U \sum_{\omega} \Big[\left(n_{\omega 1 \uparrow} n_{-\omega 1 \downarrow} + n_{\omega 2 \uparrow} n_{-\omega 2 \downarrow} \right) \\
        &\phantom{U \sum_{\omega} }+\frac{1}{2}\left(n_{\omega 1 \uparrow} n_{-\omega 2 \downarrow} + n_{\omega 2 \uparrow} n_{-\omega 1 \downarrow} \right) \\
        &\phantom{U \sum_{\omega} } + \frac{1}{4}\left(n_{\omega 1 \uparrow} n_{-\omega 2 \uparrow} + n_{\omega 1 \downarrow} n_{-\omega 2 \downarrow} \right) \Big],
\end{aligned}
\end{equation}
where $\{1,2\}$ refer to the orbital indices. The corresponding Landau functional is
\begin{equation}
    \Omega[\Delta_{\nu 1},\Delta_{\nu 2}] = \Omega_{\mathrm{imp}}[\Delta_{\nu 1},\Delta_{\nu 2}] -\frac{1}{\beta} \sum_{\nu} \left(\frac{\Delta_{\nu 1}^2}{t^2} + \frac{\Delta_{\nu 2}^2}{t^2}\right).
\end{equation}
The impurity Green's functions of the two orbitals $m=\{1,2\}$ can be analogously defined 
\begin{equation}
    G_{\nu m} = \frac{\beta}{2} \frac{\delta \Omega_{\mathrm{imp}} }{\delta \Delta_{\nu m}}.
\end{equation}
Thus, the DMFT equations become
\begin{equation} \label{eq:twoorbdmft}
    \frac{\beta}{2} \frac{\delta \Omega}{\delta \Delta_{\nu m}} = G_{\nu m}-\frac{1}{t^2}\Delta_{\nu m} = 0
\end{equation}
and the corresponding Hessian matrix is
\begin{equation}
    M_{\nu \nu'}=\frac{\beta}{2} \mqty(\frac{\delta^2\Omega}{\delta\Delta_{\nu 1}\delta\Delta_{\nu' 1}} & \frac{\delta^2\Omega}{\delta\Delta_{\nu 1}\delta\Delta_{\nu' 2}}\\
    \frac{\delta^2\Omega}{\delta\Delta_{\nu 2}\delta\Delta_{\nu' 1}} & \frac{\delta^2\Omega}{\delta\Delta_{\nu 2}\delta\Delta_{\nu' 2}}).
\end{equation}
Similarly, for the derivative of the Green's functions with respect to $\mu,U$ we get
\begin{align}
    \pdv{G_{\nu m}}{\mu} =& \sum_{\nu'} \left(\frac{G_{\nu m}}{\delta\Delta_{\nu' 1}} \pdv{\Delta_{\nu' 1}}{\mu} + \frac{G_{\nu m}}{\delta\Delta_{\nu' 2}} \pdv{\Delta_{\nu' 2}}{\mu}\right) \\ \nonumber
    &+ \left.\pdv{G_{\nu m}} {\mu}\right|_{\Delta_{ 1},\Delta_{2}}\\
    \pdv{G_{\nu m}}{U} =& \sum_{\nu'} \left(\frac{G_{\nu m}}{\delta\Delta_{\nu' 1}} \pdv{\Delta_{\nu' 1}}{U} + \frac{G_{\nu m}}{\delta\Delta_{\nu' 2}} \pdv{\Delta_{\nu' 2}}{U}\right) \\ \nonumber
    &+ \left. \pdv{G_{\nu m}}{U}\right|_{\Delta_{1},\Delta_{2}}.
\end{align}
By using the generalized susceptibilities $\chi^{\nu \nu'}_{m m'} = \chi^{\nu \nu'}_{\uparrow\uparrow m m'} + \chi^{\nu \nu'}_{\uparrow\downarrow m m'}$ and evaluating at the DMFT solution we get
\begin{align}
\label{eq:2orbital_dgdmu}
     \mqty( \pdv{G_{\nu 1}}{\mu} \\  \pdv{G_{\nu 2}}{\mu}) =& 
     \frac{1}{\beta}\sum_{\nu' m'}\left[
        \mqty(\chi^{\nu \nu'}_{11}&\chi^{\nu \nu'}_{12}\\ 
        \chi^{\nu \nu'}_{21}& \chi^{\nu \nu'}_{22})^{-1} 
        + \frac{t^2}{\beta} \mqty(\delta_{\nu \nu'}&0\\0&\delta_{\nu \nu'})
     \right]^{-1} \\ 
     \mqty( \pdv{G_{\nu 1}}{U} \\  \pdv{G_{\nu 2}}{U}) =& 
     \sum_{\nu'}\left[
        \mqty(\delta_{\nu \nu'}&0\\0&\delta_{\nu \nu'})
        +\frac{t^2}{\beta}
        \mqty(\chi^{\nu \nu'}_{11}&\chi^{\nu \nu'}_{12}\\ 
        \chi^{\nu \nu'}_{21}& \chi^{\nu \nu'}_{22})
     \right]^{-1} \\\nonumber
     &\times \left.\mqty( \pdv{G_{\nu' 1}}{U}\\ \pdv{G_{\nu' 2}}{U})\right|_{\Delta_{1},\Delta_{2}},
\end{align}
where in Eq.~(\ref{eq:2orbital_dgdmu}), we have already substituted 
\begin{equation}
    \left.\mqty( \pdv{G_{\nu 1}}{\mu}\\ \pdv{G_{\nu 2}}{\mu})\right|_{\Delta_{1},\Delta_{2}} = \frac{1}{\beta} \sum_{\nu'} \mqty(\chi^{\nu \nu'}_{11} + \chi^{\nu \nu'}_{12} \\
    \chi^{\nu \nu'}_{21} + \chi^{\nu \nu'}_{22}).
\end{equation}
Following the same steps as in the previous section, by using the Galitskii-Migdal formula \cite{Galitskii1958} for the two-orbital model
\begin{equation}
    \mathcal{D}  =  \frac{1}{\beta U} \sum_\nu \e^{\ii\nu 0^+} \left(\Sigma_{\nu1} G_{\nu1} + \Sigma_{\nu2} G_{\nu2}\right),
\end{equation}
and by re-expressing in the eigenbasis of the generalized susceptibility $\chi^{\nu\nu'}_{m m'}$ ($\sum_{\nu' m'} \chi^{\nu\nu'}_{m m'} V^\alpha_{\nu'm'} = \lambda_\alpha V^\alpha_{\nu m}$),
analog stability criteria are obtained by using the following expressions:
\begin{align}
    \pdv{n}{\mu} =& 2 \sum_{\alpha} \frac{1}{\frac{1}{\lambda_\alpha} + \frac{t^2}{\beta}} w_\alpha \\
    \pdv{n}{U} =& 2 \sum_{\alpha} \frac{1}{\frac{1}{\lambda_\alpha} + \frac{t^2}{\beta}}  y_\alpha \\
    \pdv{\mathcal{D}}{U} =& -\frac{\mathcal{D}}{U} + \sum_{\alpha} \frac{1}{\frac{1}{\lambda_\alpha} + \frac{t^2}{\beta}}  v_\alpha
\end{align}
with the corresponding weights 
\begin{align}
    w_\alpha=&\frac{1}{\beta}\sum_{\nu m}  V_{\nu m}^\alpha \frac{1}{\beta}\sum_{\nu' m'}  V_{\nu' m'}^{ \alpha-1}\\
    y_\alpha =& -\frac{1}{\beta}\sum_{\nu m} V^\alpha_{\nu m}  \frac{1}{\beta}\sum_{\nu' m'} V^{\alpha-1}_{\nu' m'} \left.\frac{\partial \Sigma_{\nu' m'}}{\partial U} \right|_{G_{1}, G_{2}}\\
    v_\alpha =& -\frac{1}{U\beta}\sum_{\nu m} (\ii\nu + \mu -2t^2 G_{\nu m}) V^\alpha_{\nu m}  \\\nonumber
    &\times \frac{1}{\beta}\sum_{\nu' m'} V^{\alpha-1}_{\nu' m'} \left.\frac{\partial \Sigma_{\nu' m'}}{\partial U} \right|_{G_{ 1}, G_{ 2}}.
\end{align}

\section{Contributions of the lowest eigenvalues to the compressibility}
\begin{figure}[t!]
\includegraphics[width=\columnwidth]{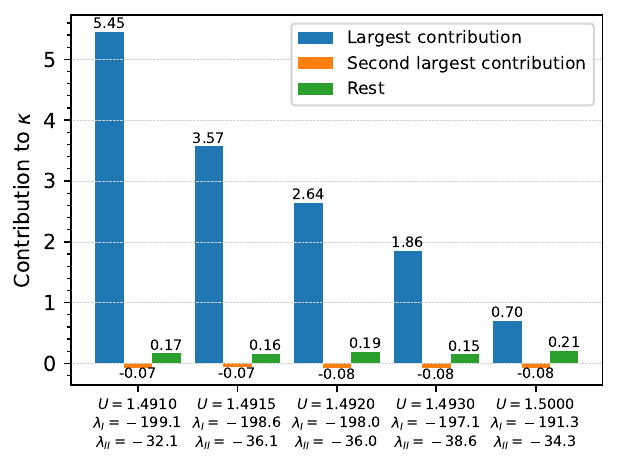}
\caption{Contributions to $\kappa$ of the lowest two eigenvalues $\lambda_I$ and $\lambda_{II}$ and the rest. We use the same interaction values as in the main panel b) of Fig.~3 in the main text, i.e., at $T_2=1/50$.}
\label{fig:contributions}
\end{figure}

Fig.~\ref{fig:contributions} shows the contributions of the lowest two eigenvalues $\lambda_I$ and $\lambda_{II}$ and the rest to  $\kappa$, given by
\begin{eqnarray}
    \kappa\!=\!\frac{2}{n^2}\sum_\alpha\left[\frac{1}{\lambda_\alpha} \! + \! \frac{ t^2}{\beta}\right]^{-1}  \kern-1em  \,
    w_\alpha,
    \label{eq:kappa}
\end{eqnarray}
calculated at the two-particle level for the two orbital model of the main text at temperature $T_2=1/50$.
One can immediately see that, approaching the divergence of $\kappa$, $\lambda_I$ gives the dominant contribution. For the single-band system, a similar trend has been reported in Ref.~\cite{Reitner2020}. The second smallest eigenvalue, on the contrary, lowers the value of $\kappa$.

\section{Data availability}
A data set containing all numerical data and plot scripts used to generate the figures for this publication is publicly available at \cite{data}.

\bibliography{references.bib}